\input{aipcheck}

\documentclass[
    ,final            
  ]
  {aipproc}

\layoutstyle{6x9}

\newcommand{\BEQ}{\begin{equation}}
\newcommand{\EEQ}{\end{equation}}
\newcommand{\BEA}{\begin{eqnarray}}
\newcommand{\EEA}{\end{eqnarray}}
\newcommand{\nn}{\nonumber}
\renewcommand{\d}{{\rm d}}

\newcommand{\omt}{{\omega_0t\,}}
\newcommand{\omtau}{{\omega_0\tau\,}}

\renewcommand{\nn}{\nonumber}
\newcommand{\eps}{\varepsilon}
\newcommand{\si}{\hat{\sigma}}

\newcommand{\om}{\omega}

\newcommand{\e}{\hat{\eta }}
\newcommand{\tr}{{\rm tr}}
\newcommand{\ha}{\hat{a}}
\newcommand{\taut}{\delta_1}

\newcommand{\T}{{\cal T}}
\renewcommand{\H}{{\cal H}}

\newcommand{\half}{\frac{1}{2}}
 
\newcommand{\Q}{{\cal{Q}}}

\def\dbarrm {{\mathchar'26\mkern-11mu{\rm d}}}                       %
\begin{document} 

\title{Bath generated work extraction 
\\ in two-level systems} 

\author{Claudia Pombo}
{address={Dept. of Visual System Analysis, AMC,
P.O.Box 12011, 1100 AA Amsterdam, The Netherlands}}
\author{Armen E. Allahverdyan}
{address={Yerevan Physics Institute,
Alikhanian Brothers St. 2, Yerevan 375036, Armenia }}
\author{Theo M. Nieuwenhuizen}
{address={Institute for Theoretical Physics, 
Valckenierstraat 65, 1018 XE Amsterdam, The Netherlands}}

\begin{abstract}
    The spin-boson model, often used in NMR and ESR physics, quantum
  optics and spintronics, is considered in a solvable limit to model a
  spin one-half particle interacting with a bosonic thermal bath.  By
  applying external pulses to a non-equilibrium initial state of the
  spin, work can be extracted from the thermalized bath. It occurs on
  the timescale $\T_2$ inherent to transversal (`quantum')
  fluctuations.  The work (partly) arises from heat given off by the
  surrounding bath, while the spin entropy remains constant during a
  pulse.  This presents a violation of the Clausius inequality and the
  Thomson formulation of the second law (cycles cost work) for the
  two-level system.
\end{abstract}

\maketitle

{\it Introduction.}  After E.L. Hahn discovered the spin-echo in NMR
physics~\cite{Hahn}, it was soon considered to be a counterexample for
irreversibility and for the second law \cite{Hahn,Waugh}; for a recent
discussion see e.g. ~\cite{Waugh,balian}.  In the present
contribution we will show in a different context that NMR-physics
contains quantum effects which should be interpreted as limits of the
second law \cite{ANnmr}.  Within the same general program we recently
analyzed the thermodynamics of the Caldeira-Leggett model for a
quantum harmonic oscillator coupled to a thermal bath~\cite{ANprl}.
At low temperatures various formulations of the second law are
violated: the Clausius inequality $\,\,\dbarrm\Q\le T\d S$ is broken, the
rates of energy dispersion and entropy production can be negative, and
certain cycles are possible where heat extracted from the bath is
fully converted into work (``perpetuum mobile''). The present
analysis of  the spin-boson model reveals a
different quantum mechanism limiting the validity of the second law.

The Hamiltonian of the problem reads:
\BEA
\label{ham}
\H&=&\H(\Delta)=\H_S+\H_B+\H_I,\quad \\ \H_S&=&
\frac{\eps}{2}\,\si _z+\frac{\Delta(t)}{2}\si_x, \quad
\H_B=\sum _k\hbar\om _k\ha^{\dagger}_k\ha _k, \quad
\H_I=\frac{1}{2}\sum _kg_k(\ha _k^{\dagger}+\ha _k)\si _z. \nn
\EEA
This is a spin-$\frac{1}{2}$ interacting with a bath of harmonic
oscillators (spin-boson model \cite{rmp,lu}); $\H_S$, $\H_B$ and
$\H_I$ stand for the Hamiltonians of the spin, the bath and their
interaction, respectively.  $\si _x$, $\si _y$ and $\si
_z=-i\si_x\si_y$ are Pauli matrices, and $\ha _k^{\dagger}$ and $\ha
_k$ are the creation and annihilation operators of the bath oscillator
with the index $k$, while the $g_k$ are the coupling constants.  For
an electron in a magnetic field $B$, $\eps= \bar g\mu_B B$ is the
energy, with $\bar g$ the gyro-magnetic factor and $\mu_B$ the Bohr
magneton.  We shall consider a situation where $\Delta(t)=0$ for almost all times.
This is a prototype of a variety of physical systems \cite{rmp}, and
known to be exactly solvable \cite{rmp,lu}, since the $z$-component of
the spin is conserved, and with it the spin energy.  Physically it
means that we restrict ourselves to times much less than $\T_1$
(relaxation time of $\si_z$).  In
ESR physics~\cite{nmr} the model represents an electron-spin
interacting with a bath of phonons, for NMR it can represent a nuclear
spin interacting with a spin bath, since in certain natural limits the
latter can be mapped to the oscillator bath.  In quantum optics it is
suitable for describing a two-level atom interacting with a photonic
bath \cite{opt}.

Starting from general physical arguments \cite{rmp}, one typically takes the
quasi-Ohmic spectral density of the bath
$J(\omega)=\sum_k g_k^2\delta(\omega_k-\omega)/(\hbar\omega_k)
=g\,\,\hbar\, \exp(-\om /\Gamma)/\pi$,
where $g$ is the dimensionless damping constant and the exponential cuts off 
the coupling at $\omega\gg \Gamma$, the maximal frequency of the bath. 
As usual, the thermodynamic limit for the bath has been taken here.

Since $\Delta=0$, one has conservation of $\si_z(t)=\si_z(0)$ 
(in the Heisenberg picture). Due to this one has ~\cite{ANnmr}:
\BEA
\label{kant}
&&\sum _kg_k[\,\ha _k^{\dagger}(t)+\ha _k(t)\,]=\e (t)-\si_z\,G(t),
\\
\e  (t)&=&
\sum_kg_k[\ha _k^{\dagger}(0)e^{i\om _kt}+\ha _k(0)e^{-i\om _kt}],
\qquad 
G(t)=g\frac{\hbar \Gamma}{\pi}\frac{\Gamma^2 t^2}{1+\Gamma^2t^2},\nn
\EEA
where $\e(t)$ is the quantum noise operator, and where the structure
of $G(t)$ shows that $1/\Gamma$ is the relaxation time of the bath.

{\it Separated initial state}.
To describe situations, where the spin was suddenly brought into the
contact with the bath, e.g. an electron injected into semiconductor,
atom injected into a cavity, or exciton created by external radiation,
we make the assumption that initially, at 
$t=0$, the spin and the bath are in a separated state,
the latter being Gibbsian at temperature $T=1/\beta$:
$\rho (0)=\rho _{S}(0)\otimes 
\exp (-\beta \H_B)/Z_B$,
where $\rho _S(0)$ is the initial density matrix of the spin.
In this situation the quantum noise is stationary and Gaussian 
with average zero. The Heisenberg equation 
$\hbar\dot{\hat{\sigma}}_\pm = i\,
\left[\,\pm \eps +\e(t)-G(t)\,\right]\,\si_\pm$ for 
$\si_\pm =\si _x\pm i\,\si _y$ can be solved exactly with the result 
\cite{ANnmr}:
\BEA
\label{k7}
\langle \si_{\pm} (t)\rangle =e^{\pm i\omt-\xi(t)}
\langle\si_{\pm} (0)\rangle,\quad
\xi (t)=\frac{g}{\pi}\,
\ln \frac{\Gamma ^2\left(1+\frac{T}{\,\,\hbar\,\Gamma}\right)
\,\,\sqrt{1+\Gamma^2t^2}}{\Gamma\left (1+\frac{T}{\,\,\hbar\,\Gamma }
-i\frac{Tt}{\hbar}\right )\Gamma \left (1+\frac{T}{\,\,\hbar\,\Gamma }
+i\frac{Tt}{\hbar}\right )},
\label{nepal}
\EEA where $\langle \e (t) \e (0)+\e (0) \e (t)
\rangle=2\hbar^2\ddot\xi(t)$, $\omega_0=\eps/\hbar$, and $\langle
\si_{x} (t)\rangle$, $\langle \si_{y} (t)\rangle$ are determined via
the real and imaginary part of $\langle \si_{\pm} (t)\rangle$.  For
$t\gg 1/\Gamma$ (\ref{nepal}) brings $\xi(t)\simeq t/\T_2$,
$\T_2=\hbar/(gT)$.  $\T_2$ can thus be identified with the transversal
decay time.

The density matrix of the spin reads
$\rho _S=\half+\half\sum_{k=x,y,z}
\langle \si _k(t)\rangle\, \si _k$.
Its von Neumann entropy equals
$S_{\rm vN}=-\tr \rho_S\ln\rho_S=-p_1\ln p_1-p_2\ln p_2$, where $
p_{1,2}=\half\pm\half|\langle{\vec{\sigma}}\rangle|$, $\vec{\sigma}=\{
\sigma_x,\sigma_y,\sigma_z\}$.
In the course of time $|\langle\vec{\sigma}(t)\rangle|$
decays to $|\langle\si_z(0)\rangle|$, which makes 
the von Neumann entropy increase.
Since there is no heat flow - the energy is conserved - this is
in agreement with a formulation of the second law: the entropy 
of closed system, or of an open system without energy transfer 
(the spin in contact with the bath), cannot decrease.

{\it A sudden pulse}.  So far we considered the Hamiltonian (\ref{ham})
with $\Delta=0$.  A fast rotation around the $x$-axis is described by
taking $\Delta\neq 0$ during a short time $\delta_1$; this is called a
fast pulse \cite{nmr}.  If $\Delta\sim 1/\taut$ is large, the evolution
operator describing the pulse becomes
$U_1=\exp(-i\delta_1\H(\Delta)/\hbar)\approx \exp({\half
  i\,\theta\si_x}) +{\cal O}(\taut)$, where
$\theta$$=$$-\delta_1\Delta/\hbar$ is the rotation angle,
$U_1^{-1}\,\si_{y\,,z}\,U_1=\si _z\sin \theta \pm\si _y\cos \theta$.
During the sudden switchings of $\Delta(t)$ from $0$ to $\Delta$ and
from $\Delta$ to $0$, the state of the system does not change, so
$\rho (t+\taut )=U_1\,\rho (t)\,U^{-1}_1$.  The work done by the
source is the change of the total energy,
\BEA
\label{2}
W_1(t)={\rm tr}\left[\rho(t)(\H(\Delta)-\H)+\rho (t+\taut)(\H-\H(\Delta))
\right]={\rm tr}\,\rho(t)(U^{-1}_1\H U_1-\H),\nn
\EEA
since $[U_1,\H(\Delta)]=0$. 

Our main interest is work extraction from the bath.
In order to ensure that the pulse does not change the energy of the spin, 
we first consider the case $\eps =0$, where the spin has no energy. 
For small $g$, $\theta=-\pi/2$ and $ t\gg 1/\Gamma$ 
the work appears to be
\BEQ\label{W1==}
 W_1=\frac{g\,\,\hbar\,\Gamma}{2\pi}+
\frac{gT}{2}\langle\si_x(0)\rangle\,e^{-t/\T_2}\EEQ 
If for a fixed $t$, temperature is neither too large nor too small,
$Te^{-t/\T_2}> \,\,\hbar\,\Gamma/\pi$, 
work can be extracted ($W_1<0$),  provided
the spin started in a coherent state  $\langle\si_x(0)\rangle= -1$. 
This possibility to {\it extract work} from the equilibrated bath
($t\Gamma\gg 1$) contradicts to the Thomson's formulation of the
second law out of equilibrium. It
disappears on  the timescale $\T_2$, because then the spin looses its
coherence, $\langle\si_{x,y}(t)\rangle\to 0$. 
Notice that any combination of $\pm\pi$ pulses (this is a 
classical variation, since the coherence is not involved)
can extract work only from a non-thermalized bath, i.e. 
for times $\sim 1/\Gamma$. Thus, the effect is indeed essentially quantum.

{\it Initial preparation via a rotation.}
Our approach also allows to consider a specific, well controllable 
non-equilibrium initial state: a Gibbsian of the total system,
$\rho_G=\exp(-\beta\H)/Z$, in which at $t=0$ the spin is rotated 
(``zeroeth pulse'') over an angle $-\half\pi$ around the $y$-axis,
$\rho(0)=U_0\,\rho_G\,U_0^{-1}$, with $ U_0=\exp(-i\pi\si_y/4)$. 
This maps $\si_x\to\si_z$, $\si_z\to-\si_x$.
Such a state models the optical excitation of the spin, as it is done
in NMR and spintronics.
Though $\rho(0)$ does not have the product form mentioned above (7), 
the problem remains exactly solvable.
Taking $\theta=-\half\pi$ one now gets 
\BEA W_1\approx\frac{g\,\,\hbar\,\Gamma}{2\pi}-
\left[\frac{\eps}{2}\sin\omega_0t+\frac{gT}{2}\cos\omega_0t\right]
\tanh\frac{\beta\eps}{2}\,\,e^{-t/\T_2}
\EEA
where $\gamma(t)=(g/\pi)\arctan[\Gamma t]$ arises from friction,
with $\gamma(\infty)=g/2$.
Typically  $g$ is small, so  work is extracted ($W_1<0$) 
when the sine function is positive. The work decomposes as 
$ W_1=\Delta U-\Delta Q$,
into the change in spin energy due to the pulse, 
$\Delta U\simeq -(\eps/2)\,
\sin\omt\tanh[\beta\eps/2]\,e^{-t/\T_2}$,
and the heat absorbed from the bath 
\BEA \Delta Q\approx\frac{g}{2}
\left[-\frac{\,\,\hbar\,\Gamma}{\pi}+T\cos\omt\,\tanh\frac{\beta\eps}{2}
\,e^{-t/\T_2}\right]
\EEA
Notice its similarity with $-W_1$ of Eq. (\ref{W1==}).
An interesting case is where work is performed by the total
system ($W_1<0$) solely due to  heat  taken from the bath ($\Delta Q>0$,
$\Delta U=0$). This process, possible by choosing $t\approx 2\pi n/\omega_0$
with integer $n$, can be considered as a cycle of a perpetuum mobile, 
forbidden by folklore minded formulations of the second law.
Indeed, under a rotation the length $|\langle\vec{\sigma}\rangle|$, 
and with it the von Neumann entropy, is left invariant, 
so one has a process with
$  \Delta Q>0$ , $\Delta S_{\rm vN}=0$,
which violates the Clausius inequality
$\Delta Q\le T\Delta S_{\rm vN}$.
The work needed at time zero to rotate the spin is
$W_0=(\eps/2)\,\tanh[\beta\eps/2]+g\,\,\hbar\,\Gamma/(2\pi)$,
representing the work done on the spin and on the bath, respectively. 
It can be verified that the total work $W_0+W_1$ is always positive, so 
Thomson's formulation for a cyclic change~\cite{ANthomson}
(here: the combination of the pulses at time $t=0$ and $t$) 
starting from equilibrium is obeyed.

{\it Two pulses in a rotated initial Gibbsian state.}
If there are many (very weakly interacting) spins, 
each in a slightly different external field, 
there appears an inhomogeneous broadening of the $\omega_0=\eps/\hbar$ 
line, for which we  assume the distribution 
\BEQ p(\omega_0)=
\frac{2}{\pi} \frac{[\T_2^\ast]^{-1}}{(\omega_0-\bar\omega_0)^2
+[\T_2^\ast]^{-2}}
\EEQ
having average $\bar\omega_0$ and inverse width $\T_2^\ast$,
typically much smaller than $ \T_2$.
In this case the gain for a single pulse is washed out, 
leaving only the loss $\Delta Q=-g\,\,\hbar\,\Gamma/2\pi$,
so two pulses are needed.
We consider again the rotated initial Gibbsian state,
and perform a first $-\half\pi$ pulse around the $x$-axis
at time $t_1$ and a second $\half\pi$ pulse 
at time $t_2=t_1+\tau$ (the standard $\half\pi$, $\pi$ 
combination would not expose an interesting role of the bath).
In the regime of small $g$ and large $t_1\sim \T_2$ 
the work in the second pulse is 
\BEA
W_2&=&\frac{g\,\,\hbar\,\Gamma}{2\pi}-\half e^{-t_1/\T_2}
\eps\sin\omtau\,\tanh\frac{\beta\eps}{2} \\
&-&\half e^{-t_2/\T_2}\tanh\frac{\beta\eps}{2}\,\cos\omega_0t_1
(\eps\sin\omega_0\tau+gT\cos\omega_0\tau)\nn
\EEA
At moderate times only slowly oscillating terms survive. They are the ones
that involve $\Delta t=t_2-2t_1$.
For the total work $W=W_1+W_2$ the averaging over $\omega_0$ brings
\BEA W&=&\frac{g\,\,\hbar\,\Gamma}{\pi} 
-\frac{\hbar}{4} e^{-t_2/\T_2}e^{-|\Delta t|/\T_2^\ast}
\tanh\frac{\beta\,\,\hbar\,\bar\omega_0}{2}\times \\
&&\left
\{\bar\omega_0\sin\bar\omega_0\Delta t\right. 
 +
[\frac{1}{\T_2}-\frac{{\rm sg}(\Delta t)}{\T_2^\ast}
(1+\frac{\beta\,\hbar\bar\omega_0}{\sinh\beta\,\hbar\bar\omega_0})]
\cos\bar\omega_0\Delta t\,\} \nn \EEA
For  $\Delta t$ near $2\pi n/\bar\omega_0$ such that the odd terms 
cancel, this again exhibits work extracted solely from the bath.

{\it Feasibility.}
Let us present several reasons favoring the feasibility of
the proposed setups: 1) Two-level systems are widespread,  because
many quantum system act as two-level systems under proper conditions; 
2) Detection
in these systems is relatively easy, since already one-time quantities 
$\langle\vec{\sigma}(t)\rangle$ completely determine the state; 
3) The harmonic oscillator bath is universal \cite{ms};
4) Work and heat were measured in NMR experiments more than 35 years ago
\cite{Schmidt}; 5) Our main effects do survive the averaging over disordered
ensembles of spins, thus allowing many-spin measurements.
6) The ongoing activity for implementation of quantum computers
provides experimentally realized examples of two-level systems, which 
have sufficiently long ${\cal T}_2$ times,
and admit external variations on times smaller than ${\cal T}_2$:
({\it i}) for atoms in optical traps ${\cal T}_2\sim 1$s, 
$1/\Gamma\sim 10^{-8}$s, and there are efficient methods for
creating non-equilibrium initial states and manipulating atoms by
external laser pulses \cite{atoms}; ({\it ii}) for an
electronic spin injected or optically excited in a semiconductor
${\cal T}_2\sim 1\,\mu$s \cite{spintronics}; 
({\it iii}) for an exciton created in a quantum dot  
${\cal T}_2\sim 10^{-9}$s \cite{exciton} (in cases ({\it ii}) and ({\it iii})
$1/\Gamma\sim 10^{-13}$s and femtosecond ($10^{-15}$s) laser pulses are
available); ({\it iv}) in NMR physics ${\cal T}_2\sim 10^{-6}-1$ s 
and the duration of pulses can be comparable with 
$1/\Gamma\sim 1 \,\mu$s.

{\it In conclusion,} we have analyzed for the spin-boson
model the validity of some non-equilibrium formulations of the second
law.  The model is relevant for almost any branch of condensed matter,
where two-level systems are described: NMR and ESR \cite{nmr},
Josephson junctions \cite{rmp}, quantum optics \cite{opt}.  Our
main finding is that quantum coherence puts definite limits on the
validity of non-equilibrium Thomson's formulation of the second law
and on the validity of Clausius inequality. The effects disappear in
the classical limit.  The detailed discussion on the feasibility of
the obtained effects can be found in \cite{nmr}.


\begin{thebibliography}{99}

\bibitem{Hahn} E.L. Hahn, Phys. Rev. {\bf 80} (1950) 580;
R.G. Brewer and E.L. Hahn, Sci. Am., {\bf 251} \# 6 (1984) 42.


\bibitem{Waugh} J.S. Waugh, in {\it Pulsed Magnetic resonance: 
NMR, ESR and Optics (A recognition of E.L. Hahn)}, 
G.G. Bagguley, ed. (Clarendon, Oxford, 1992) pp. 174.

\bibitem{balian}R. Balian, {\it From Microphysics to Macrophysics},
Vol. 2, (Springer, Berlin, 1992).

\bibitem{ANnmr}  A.E. Allahverdyan and Th.M. Nieuwenhuizen, 
cond-mat/0201408. 


\bibitem{ANprl}  A.E. Allahverdyan and Th.M. Nieuwenhuizen, 
Phys. Rev. Lett. {\bf 85}, 1799 (2000);
Phys. Rev. E {\bf 64} 056117 (2001).
Th.M. Nieuwenhuizen and  A.E. Allahverdyan, 
cond-mat/0011389.

\bibitem{ANthomson} 
A.E. Allahverdyan and Th.M. Nieuwenhuizen, Physica A {\bf 305},  542 (2002).


\bibitem{rmp} A.J. Leggett, et al., 
Rev. Mod. Phys., {\bf 59}, 1 (1987).

\bibitem{lu}J. Luczka, Physica A, {\bf 167}, 919 (1990).

\bibitem{nmr}A. Abragam, {\it Principles of Nuclear Magnetism}, 
(Clarendon, Oxford, 1961);
R.R. Ernst, G. Bodenhausen, A. Wokaun, {\it Principles of Nuclear Magnetic
Resonance in One and Two Dimensions}, (Clarendon, Oxford, 1987).



\bibitem{opt}M. Scully and S. Zubairy, {\it Quantum Optics}, (Cambridge
University Press, 1997). O. Kocharovskaya, Phys. Rep., {\bf 219}, 175,
(1992).
\bibitem{ms}
K. M\"ohring and U. Smilansky, Nucl. Phys. {\bf A338}, 227, (1980);
A.O. Caldeira and A.J. Leggett,  
{\it Quantum tunneling in a dissipative system}, 
Ann. Phys. {\bf 149}, 374-456, (1983).


\bibitem{Schmidt} J. Schmidt and I. Solomon, 
J. Appl. Phys. {\bf 37}, 3719 (1966). 


\bibitem{atoms} 
J.I. Cirac and P. Zoller, {\it 
Quantum Computations with Cold Trapped Ions},
Phys. Rev. Lett. {\bf 74}, 4091 (1995);
D. Frese, B. Ueberholz, S. Kuhr, W. Alt, D. Schrader, V. Gomer, and 
D. Meschede, {\it  Single Atoms in an Optical Dipole Trap: 
Towards a Deterministic Source of Cold Atoms}, 
Phys. Rev. Lett. {\bf 85}, 3777 (2000). 






\bibitem{spintronics}
J.M. Kikkawa and D.D. Awschalom, Science, {\bf 287}, 473 (2000).

\bibitem{exciton} N. H. Bonadeo, G. Chen, D. Gammon, D.S. Katzer, 
D. Park, and D.G. Steel, Science, {\bf 282}, 1473 (1998);
{\it Nonlinear Nano-Optics: Probing One Exciton at a Time},
Phys. Rev. Lett. {\bf 81}, 2759 (1998).

\end{thebibliography}
\end{document}